# Low- and high-temperature anomalies in the physical properties of solid methane

A.V.Leont'eva[1], A.Yu.Zakharov[2], A.Yu.Prokhorov[1,2]

[1]Donetsk Physics and Engineering Institute named after A.A.Galkin, Donetsk, 83114, Ukraine,
tonya.leont@gmail.com
[2]Yaroslav-the-Wise Novgorod State University, Veliky Novgorod, 173003, Russia,
a.yu.zakharov@gmail.com; anatoly.zakharov@novsu.ru; vesta-news@yandex.ru

Abstract

The anomalous behavior of thermodynamic, spectral, plastic, elastic and some other properties of solid methane is discussed near 20.48K and within the temperature range 60−70K. In the vicinity of 20.48 K the anomalies are due to the well-known phase α–β transition. Anomalies in a relatively wide range of 60−70 K are shown to be related to the essentially quantum nature of the collective rotation of methane molecules. The provided analysis of the high-temperature anomalies encompasses a more extensive set of phenomena in comparison with the low-temperature ones.

## 1. Introduction.

Solid methane is the lightest representative of the simplest molecular crystals (hydrocarbons) formed by tetrahedral molecules of the $CX_4$ type (symmetry **4-3m**). The methane molecule is of the type of a spherical top.

In 1953, by the NMR method it was established that near 60 K the character of the rotation of solid methane molecules from the hindered to the more free, approaching the liquid state, is sharply changed [1]. The most important thing in this discovery is that the rotational degrees of freedom of molecules in solid methane are not frozen, as one would expect.

As it is well-known, for each degree of freedom there is a characteristic temperature below which the quantum nature of matter begins to manifest itself. For vibrational degrees of freedom in a crystal (phonons), the characteristic temperature is the Debye temperature. For rotational degrees of freedom, the rotational temperature is $T_{rot} = \dfrac{\hbar^2}{2I}$ where $I$ is the moment of inertia of the top.

The qualitative difference between the dependences of the vibrational and rotational components of the heat capacity is that the vibrational heat capacity is a monotonic function of temperature, and the rotational component has a characteristic maximum at temperature $T_0 \approx 0.6\,T_{rot}$ (Fig. 1) [2]. If we start from the known value of the moment of inertia of the methane molecule, then the rotational temperature of methane should be at a temperature of the order of 7.5 K. However, analysis of the spectroscopic data of methane located on the planets of the gas giants of the solar system leads to rotational temperatures of the order of 100 K [3,4].

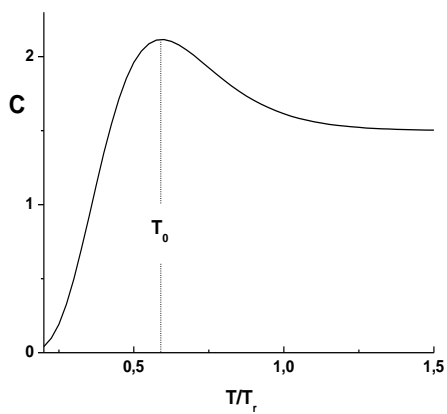

Fig. 1. Dependence of the rotational heat capacity on the dimensionless temperature in the vicinity of $T_0$ [2].

A physical explanation of this paradox was proposed in [5–7]. A model of collectivization of the rotational degrees of freedom of methane molecules was proposed and it was shown that the effective moment of inertia of the corresponding elementary excitations (topons) in solid methane depends on external conditions and is approximately 15–30 times less than the moment of inertia of a single methane molecule.

Thus, for the first time estimation was made of the temperature of the maximum of the rotational heat capacity $T_0$ and the rotational temperature $T_{rot}$ for solid methane. All earlier calculations of these temperatures before [5–7] up to Landau [2], were made only for individual molecules, i.e. for gaseous methane.

The proposed model has made it possible to explain the anomalous behavior of the specific heat of solid methane observed in the range of 60–70 K, which was observed earlier [8].

As indicated above, the quantization of rotation leads to the appearance of a maximum on the temperature dependence of the rotational heat capacity of gaseous methane (Fig. 1) [2]. In the solid state, the same effect is manifested not only in the heat capacity, but also in other thermodynamic, structural and kinetic properties, for example, NMR spectroscopy, low-frequency internal friction, adhesion, and other properties. However, in these cases it is no longer necessary to consider individual, but collective rotational disturbances.

Thus, in addition to the specific heat, anomalous behavior of a number of other physical properties of solid methane takes place in this temperature range (see below). Experimental data on these features of solid methane were first systematized in work [6].

In this paper, we perform a comparative analysis of the behavior of the properties of solid methane in the vicinity of the α–β phase transition point and in the temperature range 60 – 70K.

## 2. Phase transition in CH$_4$, $T_{\alpha-\beta} = 20.48K$

Crystalline methane at the equilibrium vapor pressure (triple point $T_{tr} = 90.67K$, fcc lattice) from the low-temperature cubic phase (space group **Pm3m** or **P4-3m**) at 20.48K transforms into an orientation-disordered in hydrogen atoms fcc phase (the space group **Fm3m**) in which the hydrogen atoms revolve around the carbon atom. That may be a sign of the manifestation of quantum effects in methane, since it occurs at low temperatures. This phase transition in methane at $T_{\alpha-\beta} = 20.48K$ was considered in detail in [10–13].

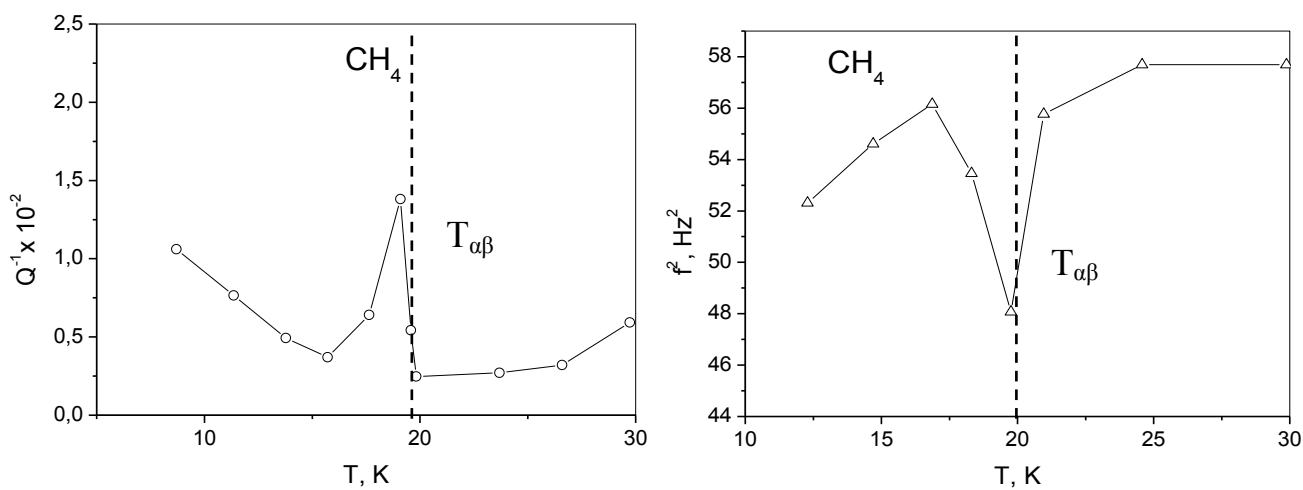

Fig. 2a. Temperature dependence of internal friction Q$^{-1}$ (T) of solid methane in the vicinity of $T_{\alpha-\beta}$.

Fig.2b. Temperature dependence of the square of the frequency of torsion oscillations f$^2$(T) in crystalline methane in the vicinity of $T_{\alpha-\beta}$.

Figures 2a and 2b show the temperature dependences of the low-frequency internal friction (LFIF) $Q^{-1}(T)$ and the square of the frequency of the torsion oscillations $f^2(T)$ in solid methane, which show its anomalous behavior just near $T_{\alpha-\beta} = 20.48K$. Namely, near 20 K, characteristic peak of $Q^{-1}(T)$ and a clear minimum on $f^2(T)$ are clear visible, which are caused by the transition of methane from the state of orientational ordering to the orientation-disordered fcc phase.

### 3. High-temperature transition between the quantum and classical regimes in solid methane (60–70 K)

It was found by Tomita [1] that in the temperature range 60–65 K, the half-width of the resonance absorption lines $H_{1/2}$ sharply decreases (Fig. 3), and the spin-lattice relaxation time $\tau_c$ (Fig. 4) sharply changes. This was interpreted as a transition to the disinhibited rotation of $CH_4$ molecules, accompanied by a twenty-fold change in the activation energy of the rotational motion of methane molecules in the temperature range 60–70 K.

However, a different interpretation of the phenomena occurring in solid methane in the indicated temperature interval was proposed in [5–7]. A model was proposed for the collectivization of the rotational degrees of freedom of methane molecules in the solid phase. This model made it possible to explain the anomalous behavior of the rotational component of the heat capacity of solid methane and to reconcile these results with spectroscopic data on the measurement of the rotational temperature of methane in the conditions of gas giants (Jupiter, Saturn, and their satellites).

**4. Comparative analysis of the anomalies of solid methane in low- and high-temperature intervals.**

Subsequently, in numerous studies, the anomalous behavior of not only the heat capacity, but also many other physical properties of solid methane in the temperature range of 60–70 K was observed. A list of the anomalies of the properties of solid methane in this interval in comparison with the corresponding anomalies in the alpha-beta phase transition is shown in Figs 3–14.

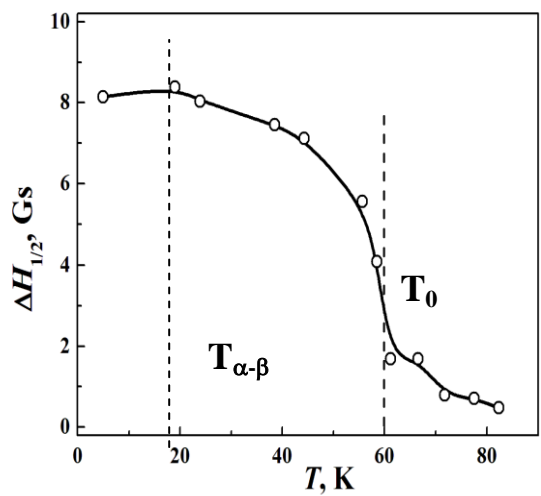
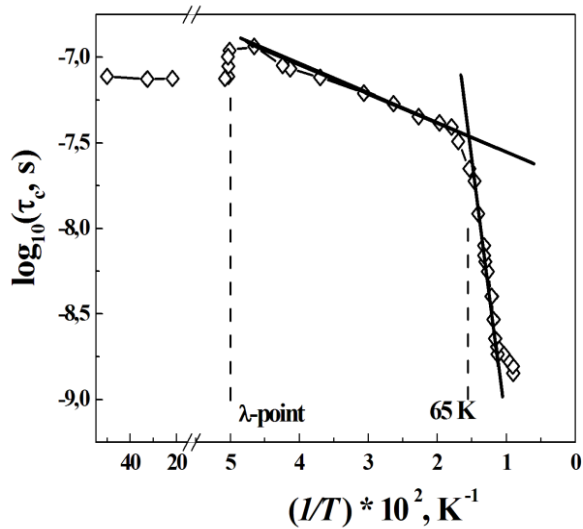

Fig.3.                                                                 Fig.4.

Fig.3. Temperature dependence of half-width of resonance absorption lines $H_{1/2}$ of the dynamical local field [1]. One can see a large jump in $\Delta H_{1/2}$ at T = 60K and a small change near 20K.

Fig.4. The dependence of characteristic time $\tau_c$ of spin-lattice relaxation on reciprocal temperature [1]. A sharp break is seen near T = 65 K.

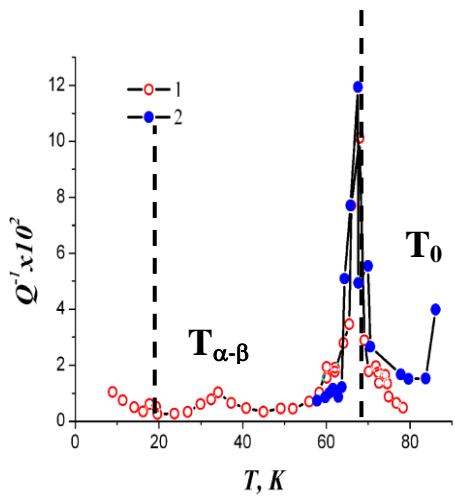 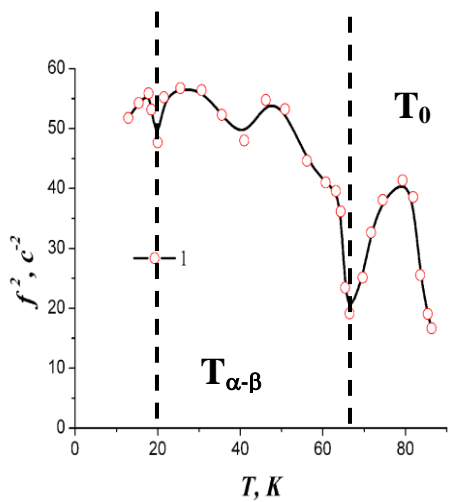

Fig.5.    Fig.6.

Fig.5. Temperature dependence of LFIF $Q^{-1}$ of solid methane [15]: 1 – cooling mode; 2 – heating mode.

Fig.6. Temperature dependence of square of torsion oscillations frequency $f^2$ of solid methane [15].

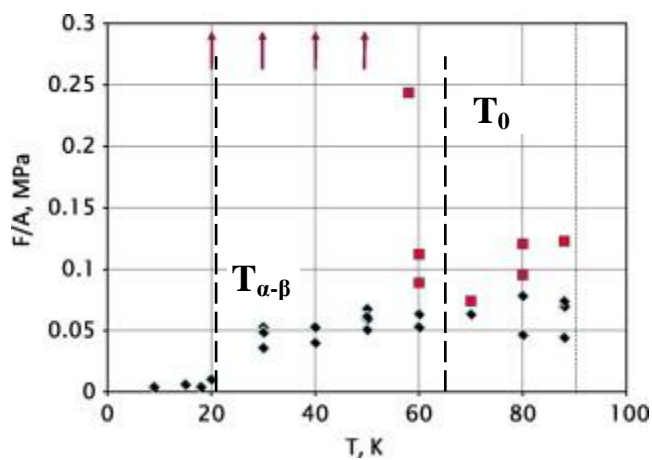 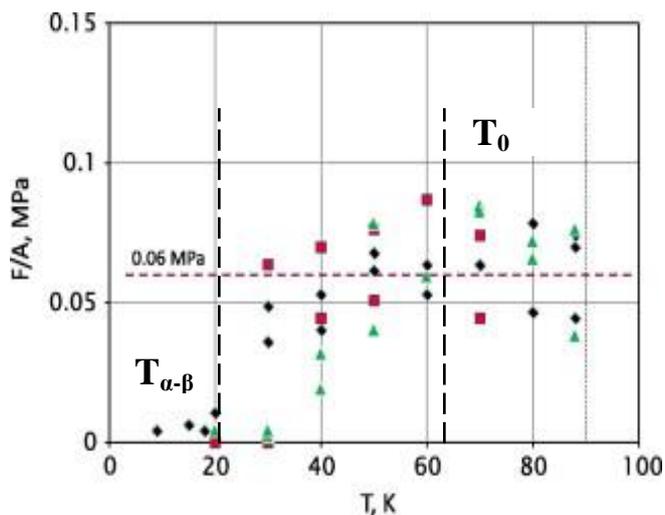

Fig.7    Fig.8.

Fig. 7. Shear stresses corresponded to breaking point of the bond between solid methane and aluminum probe for different temperatures. Square data points represent bond established at 70 K and breaking test conducted at temperature of interest. Rhomb points represent bond established and breaking test conducted at the same temperature [14].

Fig. 8. Temperature dependence of shear stresses corresponding to the breaking point of the bond between solid methane and probe for different materials of the probe: rhomb data points – aluminum alloy 7075; square data points – stainless

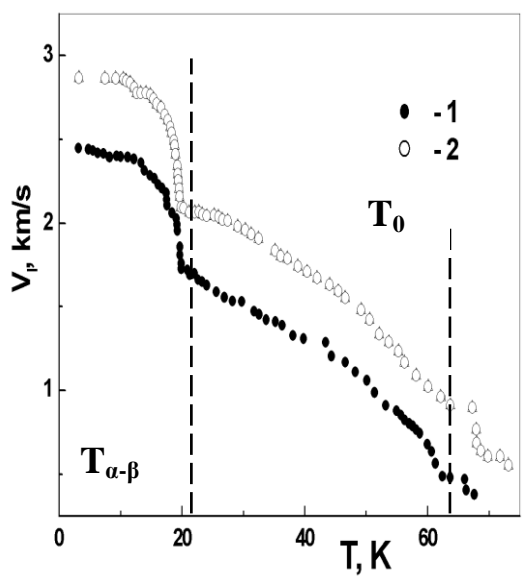 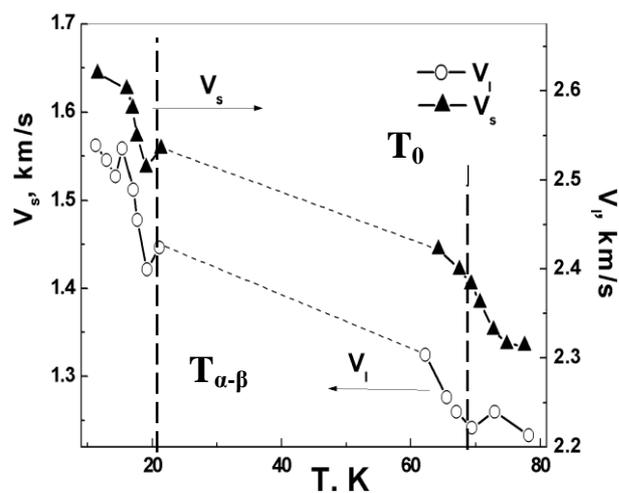

steel; triangle data points – PTFE [14].

Fig.9                                Fig.10.

Fig.9. Temperature dependence of longitudional sound velocity $V_l$ for the samples grown at various modes: 1 – fast growth; 2 – slow growth [9].

Fig.10. Temperature dependence of the velocity of longitudinal $V_l$ and transverse $V_s$ sound of solid methane in the vicinity of the phase transition temperature and in temperature interval 60–80K [6].

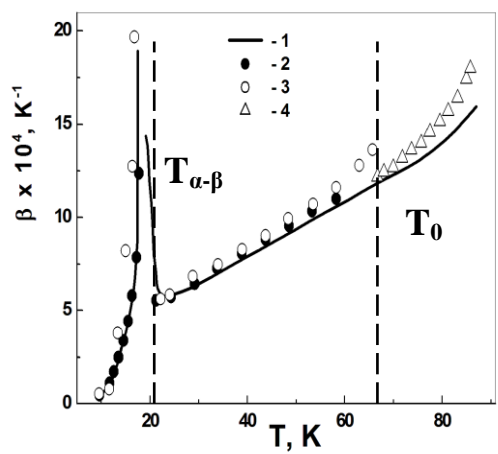 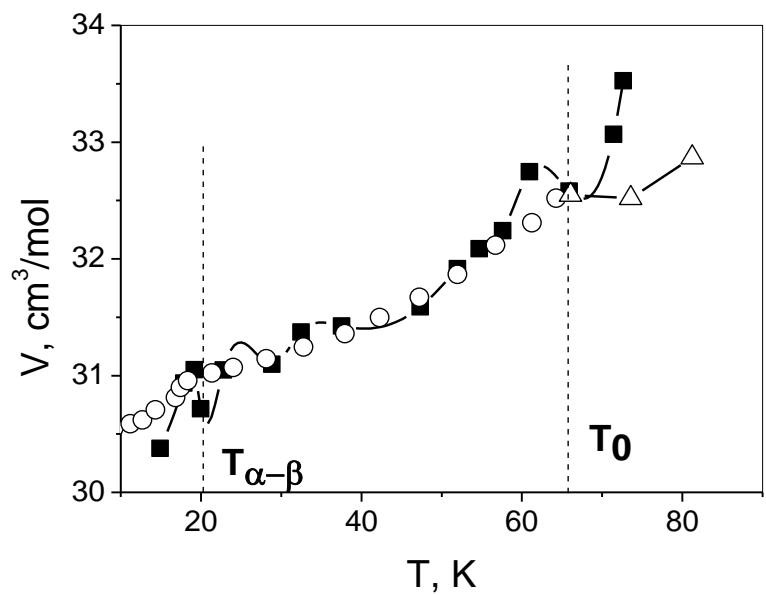

Fig.11.                                                         Fig.12.

Fig.11. Temperature dependencies of volume coefficient of thermal expansion β for solid methane according to the data of several papers [9].

Fig.12. The temperature dependencies of molar volume for solid methane according to the results of several authors [9].

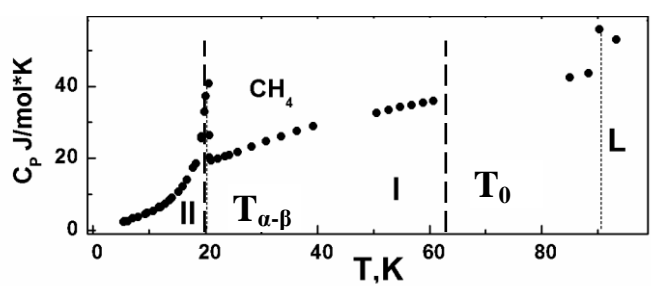 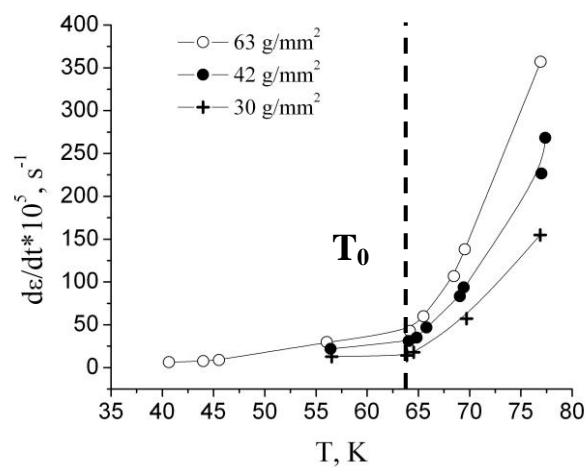

Fig.13                                                          Fig.14.

Fig. 13. Temperature dependence of heat capacity $C_P(T)$ for $CH_4$ according to [8]. I – high-temperature disordered phase, II – low temperature ordered phase, L – liquid state.

Fig.14. Temperature dependencies of creep velocity for solid methane samples under different stresses [15].

Thus, in addition to the high-temperature anomalies found by Tomita in NMR spectroscopy, there are peculiarities in the temperature dependences of many physical properties of solid methane, but especially this applies to adhesion [14] and to low-frequency internal friction (LFIF) [15] in the temperature interval 60-70K, where the anomalies are most significant.

In studying the temperature dependence of adhesion, the most important result was obtained near the temperature of 60 K, where the temperature dependence of the shear stress corresponding to the breakdown of the bond between solid methane and probes reveals a minimum (Fig. 7). It is important that the temperature of this minimum at $T = 60-70K$ does not depend on the material of the probe, by means of which adhesion was measured (see Fig.8). In other words, it reflects precisely the critical change in the physical properties of methane itself.

This result is surprisingly correlated with the data of the temperature dependences of the LFIF of solid methane (Fig. 5), where an anomalously high LFIF peak $Q^{-1}$ was detected near the same temperature ($T \sim 60K$), and also a minimum on the temperature dependence of the square of the frequency $f^2(T)$ [15].

It should be noted that in the case of LFIF, as in adhesion and NMR spectroscopy data, high-temperature anomalies are much higher than low-temperature anomalies at 20.48 K.

Surprisingly, this coincidence of the temperatures of the anomalies in these two cases takes place in different studies made by different authors.

Figure 14 shows the temperature dependence of the steady-state creep rate $\dot{\varepsilon}$ (tension at a constant load) of crystalline methane at various stresses in the range 40–80 K. One can clearly see the difference in the behavior of the dependence of $\dot{\varepsilon}$ below and above the critical temperature 65K, which was discussed above. Thus, at a stress $\sigma = 63 g/mm^2$, the value of the derivative $d\varepsilon/dT$ changes by a factor of 20 over the above-mentioned $T = 65K$ [15].

It should be noted that the features of the physical properties of solid methane, similar to those described near these temperatures, are described in detail in [5–7] on the basis of a large number of studies of various thermodynamic properties of solid methane, where anomalies missed by the authors of [9] are unambiguously identified (Figures 3–14) in the range 60–70 K. The high-temperature anomalies sometimes even exceed the anomalies near the α-β transition ([1,15]). The exception is the figure 13 [8], explicitly edited by the authors, where the temperature dependence of the heat capacity of solid methane contains an anomaly only at a low-temperature phase transition (20 K), and at high temperatures there appears to be an artificial dip just in the range 60–80 K, i.e. at temperatures of the quantum-classical transition.

## 5. Rotational temperature $T_{rot}$ – criterion of the aggregate state of methane on the planets of the solar system

In recent years, interest in crystalline methane has increased, as in the latest technologies (for example, solid methane is indispensable as a moderator in pulsed sources of cold neutrons), and also in astrophysics. The list of planets and their satellites, in the atmosphere of which methane is detected, is quite impressive:

Titan, Mars, Jupiter, Saturn and their satellites, Pluto, planets of other star systems, brown dwarfs, and also comets [16].

The increased interest in methane on planets and satellites is apparently due to the fact that methane, as is known, is formed only when organic substances of plants or living organisms decay.

As we indicated earlier, there is an independent experimental method for determining the rotational temperature: the determination of the value of $T_{rot}$ from spectroscopic data both in terrestrial conditions and on planets (and satellites of the planets) of the solar system. Thus, the rotational temperature of methane, obtained from spectroscopic data for Saturn, is between 122K and 142K. Similar measurements for Jupiter give values of the rotational temperature in the interval from 150 K to 230 K [3,4].

A comparison of the critical temperature data for the three planets is given in Table 1.

**Table 1**

| Planet | $T_0$, K | $T_{rot}$, K |
|---|---|---|
| Earth | 62 – 68[5] | 90 – 100 |
| Saturn | 84 – 98 | 122 – 142[3] |
| Jupiter | 104 – 173 | 150 – 230 [4] |

The discrepancies between the rotational temperatures of methane for the Earth, Saturn and Jupiter are probably related to the difference in physical conditions on these planets. Hence follows the hypothesis that gives an answer to the question of the aggregate state of methane on Saturn and Jupiter. Most researchers are inclined to believe that methane, found spectroscopically on these planets, is in the gaseous state. Such a conclusion seems questionable, since the rotational temperature of the gas is determined solely by the moment of inertia of the molecules, i.e. for gaseous methane it must be everywhere the same (of the order of 7K) and not depend on external conditions.

The condensed state of methane, the rotational temperature of which is determined by the effective moment of inertia of the quasiparticles (topons), which depends on external conditions, is another matter. This fact of the dependence of $T_{rot}$ on external conditions only for the condensed state indicates that on the planets of gas giants there is a condensed phase of methane and namely its contribution is the determining one.

Thus, the high values of the rotational temperatures of methane on Jupiter and Saturn [3,4], as well as their differences on these planets, support the hypothesis that methane on these planets is in a condensed state. This hypothesis is also supported by the fact that the rotational temperature of methane on Saturn, and especially on Jupiter, lies within fairly wide limits – about 20K and 80K, respectively, which means a wide spread of thermodynamic conditions of methane in various parts of these giant planets.

## 6. Is methane liquid-like at temperatures above 65 K?

As already mentioned, Tomita [1] expressed the hypothesis of a sharp freedom of the rotation of methane molecules in the temperature range 60–70 K, which approximates methane state to the liquid one. However, experiments on the plasticity of solid methane indicate another. Even at 77K, which is much higher than $T_0$, methane still remains a crystal, but with a very high plasticity, much higher than its plasticity below 60K.

Figures 15 and 16 show the types of fractures during the tensile of methane samples up to failure. A brittle fracture is seen at temperatures below 7K and viscous – at T above $T_0 \cong 64 K$, when the failure of the sample is preceded by the formation of local thinning, the so-called "neck", which is preceded by a very high tensile strain exceeding 75%, i.e. the so-called superplasticity [17].

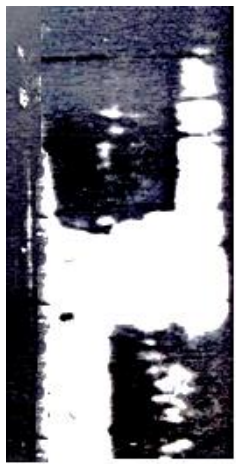 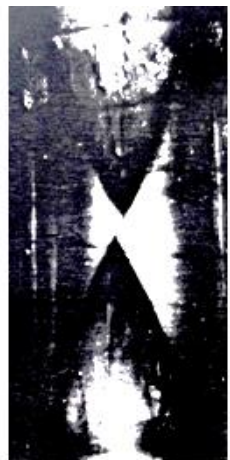

Fig.15          Fig.16

Fig.15. Photo of brittle failure of a methane sample at temperatures below 7K.

Fig.16. Photo of the plastic failure of a sample of solid methane at T above 64 K with the formation of local thinning (so-called "neck").

It should be noted that the hermetic ampoule with solid methane, the fractures of which are represented in photos 15 and 16, is surrounded by special glass chambers and immersed in a Dewar with liquid nitrogen, through the lumens in which the survey was performed [18].

Thus, the superplasticity appears above the temperature of the quantum-classical transition $T_0$ in methane, which indicates rather a significant weakening of the bonding forces (the tensile strength at $T=77K$ is only $\sigma=0.025\,Kg/mm^2$, whereas, for example, at $T=4K (T<T_0)$, it is $\sigma=0.25\,Kg/mm^2$) than about the approach of methane to the liquid state. Therefore, the above experiments do not support the Tomita's idea about liquid-like state of solid methane at T above 65K. Rather, these experimental results confirm the proposed by us model of the quantum-classical transition in methane in temperature range 60-70K.

### 7. Conclusions.

1. Crystalline methane has two characteristic temperatures at which the behavior of the molecules changes: a) phase transition at 20.48 K; b) the transition between the quantum and classical regimes of the collectivized rotational degrees of freedom of methane molecules at temperatures in the range of 60–70 K, which is manifested in NMR spectroscopy, heat capacity, low-frequency internal friction, adhesion, etc.
2. On the basis of experimental data on the plasticity and the nature of the failure of solid methane samples, it was concluded that there is a superplasticity phenomenon of methane at temperatures above 60 K, possibly due to the transition from the quantum to the classical regime of the collectivized rotational degrees of freedom of methane molecules.
3. It should be noted that in all works on the rotational heat capacity, up to Landau–Lifshitz book [2], only free methane molecules, i.e. ideal gas, were considered. For the first time for crystalline methane, the rotational temperature was calculated only in our works [5–7]. It turned out to be much higher than for a gas ($T_{rot}=7K$ for a gas and more than 100K for a crystal).
4. The agreement between the estimates of the rotational temperature of crystalline methane with the values of the rotational temperature obtained by other independent methods for the planets of the solar system supports the model of the transition between the quantum and classical regimes of the rotational degrees of freedom of crystalline methane. This result allows us to draw conclusions about the aggregate state of methane on gas giants using the spectroscopic data analysis. It is a new method for cosmic problems.

### 8. Acknowledgements.

We wish to thank L. P. Pitaevsky and O. I. Kirichek for useful discussion of the work.

*Two of the authors (A. Yu. Zakharov and A. Yu. Prokhorov) are grateful to the Ministry of Education and Science of Russian Federation for financial support by within the framework of the project part of the state order (Project No. 3.3572.2017).*